\documentclass[cameraready]{Interspeech}

\title{SDP-Codec: A Speaker-Decoupled Speech Codec with Pitch Injection for Low-Bitrate Coding and Zero-Shot Voice Conversion}

\author[affiliation={1}, orcid=0009-0003-0635-0124]{Hounsu}{Kim}
\author[affiliation={1}, orcid=0000-0003-2664-2119, correspondingauthor]{Juhan}{Nam}

\address{
    $^1$ Graduate School of Culture Technology, KAIST, Daejeon, South Korea
}

\email{hanshounsu@kaist.ac.kr, juhan.nam@kaist.ac.kr}

\keywords{Neural speech codec, discrete speech tokens, low-bitrate coding, speaker decoupling, speech language modeling}

\usepackage{comment}
\usepackage{xurl}  
\urldef\huberturl\url{https://huggingface.co/facebook/hubert-large-ls960-ft}
\usepackage{makecell} 
\usepackage{multirow} 
\usepackage{arydshln, amssymb} 
\usepackage{pifont}
\newcommand{\cmark}{\ding{51}}
\newcommand{\xmark}{\ding{55}}

\begin{document}

\maketitle

\begin{abstract}
    Speaker-decoupled speech codecs can reduce bitrate by separating global speaker attributes from local content and prosody, while supporting voice conversion.
    Existing speaker-decoupled codecs face a trade-off: methods that explicitly suppress speaker leakage often rely on multi-stage or auxiliary training, whereas simpler designs can leave residual speaker information in local tokens.
    We propose SDP-Codec, a speaker-decoupled, pitch-injected codec trained with a single-stage optimization pipeline.
    SDP-Codec derives local tokens from continuous pre-quantization features of a pretrained self-supervised encoder and injects normalized F0 via a pitch encoder-decoder with global-conditioned denormalization and soft-label pitch reconstruction objective.
    Across 16\,kHz and 24\,kHz settings, SDP-Codec achieves competitive reconstruction and strong zero-shot voice conversion at comparable bitrates, with the lowest speaker-probing accuracy among compared systems, suggesting reduced speaker leakage.
    \footnote{Source code and demo: github.com/hanshounsu/sdpcodec-open/}
\end{abstract}

\section{Introduction}
\label{sec:intro}

Neural speech codecs~\cite{zeghidour2022soundstream, defossez2023encodec} convert speech waveforms into discrete token sequences and have become a core foundation for speech language models (SLM)~\cite{wang2023valle, du2024cosyvoice2}.
Reducing bitrate is central to codec design and also benefits downstream SLMs by lowering the cost of autoregressive prediction~\cite{guo2024lscodec, zheng2025say, li2025flexicodec}.
One principled way to reduce bitrate is speaker decoupling, which decomposes speech into global, time-invariant attributes such as speaker timbre and local, time-varying attributes such as content and prosody.
By compressing only the local branch and conveying speaker identity through a separate global branch, a codec can reduce bitrate while preserving reconstruction quality.
This global/local factorization is widely used in voice conversion (VC)~\cite{qian2019autovc, choi2021neural, zhang2025vevo}, text-to-speech (TTS)~\cite{du2024cosyvoice2, wang2025sparktts}, and neural speech codec design~\cite{guo2024lscodec, ren2024ticodec, li24singlecodec, ju2024naturalspeech, zheng2025freecodec}.

Other low-bitrate strategies have also been explored, but each introduces additional assumptions or constraints.
Text-aligned tokenizers~\cite{wang2025tadicodec, tseng2025taste} produce compact representations by leveraging text supervision, but require text at inference time.
Variable-frame-rate methods~\cite{zheng2025say, li2025flexicodec, wang2025codecslime} reduce token rates by allocating fewer tokens to acoustically redundant regions, but they require an additional duration predictor.
VC-oriented methods often impose a severe bottleneck on content tokens and rely on flow-matching-based decoders for reconstruction~\cite{zhang2025vevo, joglekar2025ezvc}; although such systems can also be viewed as low-bitrate codecs, their decoders are generative rather than faithfully reproducing the input waveform.
Speaker decoupling avoids these constraints and enables compact local tokenization, direct waveform reconstruction, and inherent VC capability.

Achieving clean speaker disentanglement in neural codecs remains challenging.
Existing approaches often rely on gradient reversal~\cite{ju2024naturalspeech} or perturbation~\cite{guo2024lscodec, choi2021neural}, which increase training complexity and may introduce instability.
LSCodec~\cite{guo2024lscodec}, an early speaker-decoupled single-codebook low-bitrate codec, uses a three-stage training pipeline with speaker perturbation.
BiCodec~\cite{wang2025sparktts} also uses speaker decoupling, but does not impose explicit disentanglement constraints, which can allow speaker information to leak into local tokens.

To address these limitations, we propose \textbf{SDP-Codec}, a \textbf{S}peaker-\textbf{d}ecoupled, \textbf{p}itch-injected, single-codebook, low-bitrate codec trained in a single stage.
The local branch builds on a pretrained vq-wav2vec encoder~\cite{baevski2019vq}, whose discrete units $\hat{\mathcal{Z}}$ are widely used in VC as content representations~\cite{guo2024vec2wav}.
However, these units can be too coarse to preserve fine-grained content and prosodic details.
SDP-Codec instead re-quantizes the continuous pre-quantization features~$\mathcal{Z}$ rather than $\hat{\mathcal{Z}}$, which better preserve content but also carry more low-level information, namely global speaker timbre and local prosody.
A compact codebook then imposes a strong bottleneck that strips both types of low-level information from the local stream.
Since the local branch should still contain prosody, we reinject speaker-normalized F0 through a pitch encoder-decoder with a soft-label pitch reconstruction loss~\cite{luo2025fcpe}, while the global branch models speaker timbre and also recovers speaker-dependent pitch range.
SDP-Codec achieves competitive 24\,kHz reconstruction quality, the lowest speaker-probing accuracy, and strong zero-shot VC performance in speaker similarity, F0 correlation, and MOS at comparable bitrates.

Our contributions are as follows:
\textbf{(1)} we present SDP-Codec, a speaker-decoupled, single-codebook codec trained with a single-stage pipeline for low-bitrate speech coding;
\textbf{(2)} we introduce explicit F0 injection to inject prosody information while reducing speaker leakage in the local stream;
\textbf{(3)} we show that continuous pre-quantization vq-wav2vec features with a soft-label pitch reconstruction loss improve content fidelity in our ablations; and
\textbf{(4)} we demonstrate competitive reconstruction and strong zero-shot VC performance at comparable bitrates.

\section{Method}
\label{sec:method}

\begin{figure}[t]
  \centering
  \includegraphics[width=0.99\columnwidth]{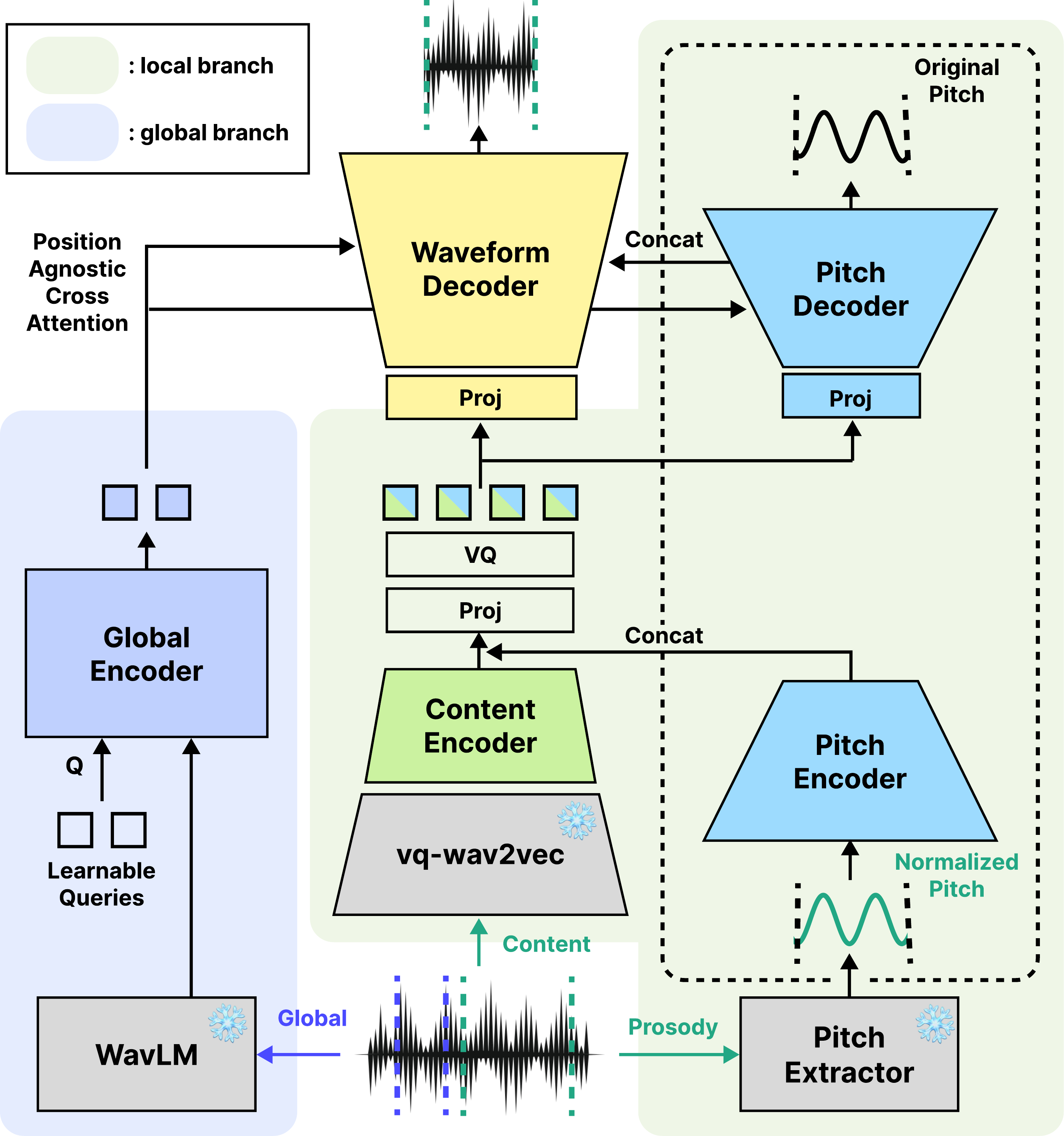}
  \caption{SDP-Codec model architecture.}
  \label{fig:model_architecture}
  \vspace{-0.75pt}
\end{figure}

As shown in Figure~\ref{fig:model_architecture}, SDP-Codec comprises a local branch and a global branch.
All pretrained components---the vq-wav2vec encoder, WavLM feature extractor, and FCPE pitch extractor~\cite{luo2025fcpe}---are frozen during training.
The local branch fuses outputs from a content encoder and a pitch encoder, jointly quantizes them with a single codebook, and decodes the resulting token stream into a waveform and F0 via a waveform decoder and a pitch decoder.
The global branch supplies time-invariant speaker embeddings to both the waveform decoder and pitch decoder.

\subsection{Content encoder}
The content encoder is built on pretrained vq-wav2vec~\cite{baevski2019vq}, whose fully convolutional structure and low speaker leakage have been shown effective in the VC literature~\cite{guo2024vec2wav}.
Rather than using the quantized units~$\hat{\mathcal{Z}}$ directly, we use the continuous pre-quantization features~$\mathcal{Z}$, which retain richer content detail.
A post-encoder of residual CNN blocks with snake activations~\cite{xin2024bigcodec} further downsamples these features to produce the content feature.

\begin{table*}[!t]
  \centering
  \footnotesize 
  \setlength{\tabcolsep}{4.2pt}
  \caption{Objective metrics for reconstruction and zero-shot voice conversion. Evaluation on LibriTTS-test-clean (24\,kHz) and LibriSpeech-test-clean (16\,kHz). WER from HuBERT CTC. Abbreviations: LT=LibriTTS, LS=LibriSpeech, Em=Emilia, MLS=Multilingual LibriSpeech, TS=TextrolSpeech, Vox=VoxCeleb, IN5=5 Indic languages.}
  \label{tab:main_results}
  \vspace{-0.75pt}
  {\renewcommand{\arraystretch}{1.1}
  \begin{tabular}{@{}l c c c c c c c c c c c@{}}
  \toprule
  Method & \makecell{VC \\ Capable} & Codec & SR & Bitrate & \makecell{Train \\ Set (hours)} &   \makecell{Test \\ Set} &
  UTMOS $\uparrow$ &
  \makecell{SECS $\uparrow$} &
  \makecell{WER $\downarrow$} &
  F0 corr. $\uparrow$ &
  STOI $\uparrow$ \\

  \midrule
  \multicolumn{12}{l}{\textbf{Reconstruction (24k)}} \\
  \midrule
  DualCodec~\cite{li2025dualcodec} (25Hz) & $\text{\xmark}$ & $\text{\cmark}$ & 24k & 0.60 & Em (100k) & LT &
  3.9503 & 0.8997 & 3.12 & 0.6537 & 0.8887 \\
  VARSTok~\cite{zheng2025say} & $\text{\xmark}$ & $\text{\cmark}$ & 24k & 0.43 & LT & LT &
  3.6498 & 0.8917 & 15.18 & 0.6097 & 0.8552 \\
  \hdashline
  LSCodec~\cite{guo2024lscodec} & $\text{\cmark}$ & $\text{\cmark}$ & 24k & 0.45 & LT & LT &
  \textbf{4.0629} & 0.9356 & 5.71 & 0.6245 & 0.7511 \\
  \textbf{SDP-Codec-24-S} & $\text{\cmark}$ & $\text{\cmark}$ & 24k & 0.45 & LT & LT &
  \underline{4.0542} & 0.9353 &
  \textbf{5.54} & \textbf{0.6522} & \textbf{0.8798} \\

  \midrule
  \multicolumn{12}{l}{\textbf{Zero-shot voice conversion (24k)}} \\
  \midrule
  vec2wav2.0~\cite{guo2024vec2wav} & $\text{\cmark}$ & $\text{\xmark}$ & 24k & 1.66 & LT & LT & 3.9349 & 0.8080 & 6.45 & 0.6143 & ---  \\
  Vevo~\cite{zhang2025vevo} & $\text{\cmark}$ & $\text{\xmark}$ & 24k & 0.65 & Internal (60k) & LT &
  3.6749 & 0.7887 & 9.66 & 0.5002 & --- \\
  \hdashline
  LSCodec~\cite{guo2024lscodec} & $\text{\cmark}$ & $\text{\cmark}$ & 24k & 0.45 & LT & LT &
  3.9034 & 0.7965 & 7.48 & 0.5357 & --- \\
  \textbf{SDP-Codec-24-S} & $\text{\cmark}$ & $\text{\cmark}$ & 24k & 0.45 & LT & LT &
  \textbf{4.0055} & \textbf{0.8133} &
  \textbf{7.04} & \textbf{0.6162} & --- \\

  \specialrule{0.08em}{0.5ex}{0.5ex}  
  \multicolumn{12}{l}{\textbf{Reconstruction (16k)}} \\
  \midrule
  FocalCodec~\cite{della_libera2025focalcodec} (50Hz) & $\text{\xmark}$ & $\text{\cmark}$ & 16k & 0.65 & LT & LS &
  \textbf{4.0591} & 0.8947 & \textbf{1.43} & {0.6287} & {0.8602} \\
  XCodec~\cite{ye2025codecdoesmatter} & $\text{\xmark}$ & $\text{\cmark}$ & 16k & 0.5 & LS & LS &
  3.8416 & 0.8487 & 3.00 & 0.6218 & 0.8354 \\
  \textbf{SDP-Codec-16-S} & $\text{\cmark}$ & $\text{\cmark}$ & 16k & 0.45 & LS & LS &
  \underline{4.0124} & \textbf{0.9372} & 4.44 & \textbf{0.6643} & \textbf{0.8724} \\
  \hline
  FlexiCodec~\cite{li2025flexicodec} & $\text{\xmark}$ & $\text{\cmark}$ & 16k & 0.52 & LL (52k) & LS &
  4.0834 & 0.9249 & 2.49 & 0.6519 & 0.8842 \\
  \hdashline
  BiCodec~\cite{wang2025sparktts} & $\text{\cmark}$ & $\text{\cmark}$ & 16k & 0.65 & Em+LS (3k) & LS &
  \textbf{4.1853} & 0.9171 & 1.98 & \textbf{0.6879} & \textbf{0.9220} \\
  MSRCodec~\cite{li2025msrcodec} & $\text{\cmark}$ & $\text{\cmark}$ & 16k & 0.52 & MLS+TS+Vox (53.6k) & LS &
  4.1382 & {0.9392} & \textbf{1.69} & {0.6366} & 0.8876 \\
  \textbf{SDP-Codec-16-L} & $\text{\cmark}$ & $\text{\cmark}$ & 16k & 0.52 & MLS+LS (45.5k) & LS &
  3.9954 & \textbf{0.9436} & 3.08 & \underline{0.6690} & \underline{0.8881} \\

  \midrule
  \multicolumn{12}{l}{\textbf{Zero-shot voice conversion (16k)}} \\
  \midrule
  EZ-VC~\cite{joglekar2025ezvc} & $\text{\cmark}$ & $\text{\xmark}$ & 16k & 0.45 & En(3.1k)+IN5(9.8k) & LS &
  4.0557 & 0.6871 & 14.18 & 0.2041 & --- \\
  \hdashline
  BiCodec~\cite{wang2025sparktts} & $\text{\cmark}$ & $\text{\cmark}$ & 16k & 0.65 & Em+LS (3k) & LS &
  3.9103 & 0.7577 & {3.43} & 0.5517 & --- \\
  MSRCodec~\cite{li2025msrcodec} & $\text{\cmark}$ & $\text{\cmark}$ & 16k & 0.52 & MLS+TS+Vox (53.6k) & LS &
  3.8966 & {0.7396} & \textbf{2.18} & {0.5806} & --- \\
  \textbf{SDP-Codec-16-L} & $\text{\cmark}$ & $\text{\cmark}$ & 16k & 0.52 & MLS+LS (45.5k) & LS &
  \textbf{3.9832} & \textbf{0.8405} & {4.11} & \textbf{0.6088} & --- \\
  \bottomrule
  \end{tabular}}
  \vspace{-0.75pt}
\end{table*}

\subsection{Pitch injection to the local stream}
We first extract a frame-wise log-F0 contour using a pretrained pitch extractor~\cite{luo2025fcpe} and apply per-segment mean-and-variance normalization to remove speaker-dependent pitch range.
Unvoiced frames are assigned a fixed value of $-3$, which falls outside the normalized pitch range.
The normalized F0 contour is fed to the pitch encoder, producing a compressed pitch feature that is concatenated with the content encoder output.
The joint feature is projected once and then quantized into a compact single codebook; the small codebook capacity (300 or 1{,}536 entries; see Sec.~\ref{sec:impl}) acts as a tight information bottleneck, constraining the local tokens to encode high-level content information while discarding low-level information.
After quantization, separate projection layers map the discrete features to the waveform decoder and pitch decoder for waveform reconstruction and original (denormalized) F0 reconstruction, respectively.

Our design departs from~\cite{ye2025codecdoesmatter} in three ways: (1)~the auxiliary reconstruction target is pitch (in 360 cent bins) rather than semantic features; (2)~pitch-decoder hidden states are concatenated with waveform-decoder hidden states at each layer to reinforce F0 in the reconstructed waveform; (3)~the pitch decoder is conditioned on global speaker embeddings to recover the mean and variance of the log-F0 contour, delegating the speaker-dependent pitch range to the global branch.

The pitch encoder applies a single downsampling step, as the input F0 contour is already near the target frame rate; it shares the content encoder's block structure with LeakyReLU in place of snake activations.
At each waveform decoder layer~$l$, let $h^w_l \in \mathbb{R}^{N \times d^w}$ and $h^f_l \in \mathbb{R}^{N \times d^f}$ denote the waveform-decoder and pitch-decoder hidden states, respectively.
We interpolate~$h^f_l$ to match the waveform-decoder resolution and concatenate: $h^w_l \leftarrow [h^w_l;\, h^f_l] \in \mathbb{R}^{N \times (d^w + d^f)}$.
Specific frame rates and encoder/decoder strides are given in Sec.~\ref{sec:impl}.

\subsection{Global branch}
The global branch extracts WavLM features~\cite{chen2022wavlm}, compresses them into fixed-length, time-invariant embeddings via a perceiver resampler~\cite{alayrac2022flamingo}, and feeds these embeddings to both the waveform and pitch decoders through position-agnostic cross-attention~\cite{du2024unicats}.
In parallel, a temporally pooled speaker embedding derived from these features conditions the adaptive snake modules~\cite{guo2024vec2wav}.
Together, these mechanisms suppress local information from the global branch and steer toward modeling only the speaker identity.

\subsection{Training Objective}

The training objective comprises four terms: (1) a multi-scale mel-spectrogram L1 loss between the input and reconstructed audio; (2) a commitment loss with the straight-through estimator for the joint content--F0 codebook; (3) an adversarial loss using a multi-scale time-domain discriminator with LSGAN~\cite{mao2017lsgan}, together with an L1 feature-matching loss; and (4) a soft-label pitch reconstruction loss~\cite{luo2025fcpe}.
The pitch decoder predicts a 360-bin cent histogram at each frame; soft targets are formed by Gaussian-blurring a one-hot bin derived from the ground-truth log-F0 contour, and binary cross-entropy is applied between these soft targets and the predicted histogram, providing smooth gradients around the true pitch.

\section{Experiments}
\label{sec:experiments}

\subsection{Datasets and Training}

We report three trained variants of SDP-Codec.
\textbf{SDP-Codec-16-S} and \textbf{SDP-Codec-24-S} are small models trained on LibriSpeech~\cite{panayotov2015librispeech} (16\,kHz) and LibriTTS~\cite{zen2019libritts} (24\,kHz) respectively, with 3.36\,s input segments.
\textbf{SDP-Codec-16-L} is a large-scale variant that adds the English subset of Multilingual LibriSpeech (MLS)~\cite{pratap2020mls} and extends the segment length to 6\,s.
We provide only a 16\,kHz large-scale variant due to resource constraints, and as most comparable baselines are evaluated at 16\,kHz.
All models are trained for 600k steps---the small models on 4$\times$RTX 4090 and the large model on 4$\times$RTX 5090---with effective batch sizes of 24 and 16, respectively, using fewer resources than typical baselines.
All models have 406\,M parameters in total, of which 74\,M are trainable.

\subsection{Baselines}

Since baselines vary in training data, bitrate, and evaluation protocol, strict comparison is inherently difficult; nevertheless, we gather recent open-source low-bitrate neural codecs and zero-shot VC systems to provide the broadest practical reference.

For SDP-Codec-16-S, we use XCodec~\cite{ye2025codecdoesmatter} and FocalCodec~\cite{della_libera2025focalcodec} as reconstruction references; for SDP-Codec-16-L, we compare against BiCodec~\cite{wang2025sparktts}, EZ-VC~\cite{joglekar2025ezvc}, FlexiCodec~\cite{li2025flexicodec}, and MSRCodec~\cite{li2025msrcodec} for both reconstruction and VC.
Of these, MSRCodec and BiCodec are the fairest comparisons as they are low-bitrate codecs targeting disentanglement.

At 24\,kHz (SDP-Codec-24-S), we compare with DualCodec~\cite{li2025dualcodec}, VARSTok~\cite{zheng2025say}, vec2wav2.0~\cite{guo2024vec2wav}, and LSCodec~\cite{guo2024lscodec}, where LSCodec is the closest prior work in both design and bitrate.
Vevo~\cite{zhang2025vevo} is included as a VC reference rather than a direct baseline given its larger training data and model scale.
For fair comparison, each baseline is configured to approximate SDP-Codec's bitrate: FlexiCodec with merging threshold $\tau=0.84$, VARSTok with similarity threshold $\tau=0.7$ and maximum cluster span $S_{\max}=4$, and DualCodec at 25\,Hz with two quantizer streams.

\subsection{Implementation Details}
\label{sec:impl}

For the joint content--F0 codebook, we use 1{,}536 entries for the large-scale model, corresponding to 0.52\,kbps bitrate, and 300 entries for the small-scale models to match 0.45\,kbps bitrate.\\
\textbf{Content (SSL) encoder.} The vq-wav2vec encoder operates on 16\,kHz waveform and produces frame-level embeddings at 100\,Hz.
The post-encoder downsamples by a factor of 2 to reach the 50\,Hz target rate.
The waveform decoder upsamples in four stages with strides [2, 4, 5, 8] for 16\,kHz output and [3, 4, 5, 8] for 24\,kHz output.\\
\textbf{Pitch encoder.} The pitch extractor outputs an F0 contour at 100\,Hz for 16\,kHz input; for 24\,kHz input, the contour is linearly interpolated to 150\,Hz.
The pitch encoder downsamples once at the final layer, yielding strides [1, 1, 1, 2] for 16\,kHz and [1, 1, 1, 3] for 24\,kHz.
The pitch decoder mirrors the encoder structure.

\subsection{Evaluation}

\textbf{Test sets.} We use LibriSpeech test-clean for the 16\,kHz experiments and LibriTTS test-clean for the 24\,kHz experiments.
For voice conversion evaluation, we randomly assign a different target speaker for each source utterance and provide a reference utterance from the target speaker for global conditioning.\\
\textbf{Objective metrics.} We report UTMOS~\cite{saeki2022utmos} for perceptual naturalness and STOI for intelligibility. We omit PESQ since it was standardized for telephony-oriented speech quality assessment and can be less reliable for modern neural codec artifacts.
Speaker identity is measured by cosine similarity (SECS) using Resemblyzer\footnote{https://github.com/resemble-ai/Resemblyzer}; content fidelity is measured by word error rate (WER) using a HuBERT-based CTC ASR model~\cite{hsu2021hubert}\footnote{https://huggingface.co/facebook/hubert-large-ls960-ft} fine-tuned on LibriSpeech.
F0 correlation is computed as the Pearson correlation of Harvest-extracted~\cite{morise2017harvest} F0 contours over voiced frames.
Objective results are summarized in Table~\ref{tab:main_results}.\\
\textbf{Subjective evaluation.} We conduct MOS tests for VC: \textit{naturalness} (NMOS) on a 5-point scale following the Absolute Category Rating (ACR) method~\cite{itu1996p800} for utterances longer than 3\,s, and \textit{speaker similarity} (SMOS) on a 4-point scale with a target-speaker reference.
We collected ratings from 27 listeners across 30 random samples for the 16\,kHz and 24\,kHz models, and report the mean and 95\% confidence interval per system.
Subjective VC MOS results are summarized in Table~\ref{tab:vc-mos}.

\begin{table}[t!]
  \centering
  \footnotesize
  \caption{Subjective voice conversion MOS results (mean $\pm$ 95\%\,CI).
  Results are grouped by sample rate (SR).}
  \label{tab:vc-mos}
  \vspace{-0.75pt}
  {\renewcommand{\arraystretch}{1.1}
  \begin{tabular}{@{}c l c c@{}}
  \toprule
  SR & Method & NMOS $\uparrow$ & SMOS $\uparrow$ \\
  \midrule
  \multirow{4}{*}{24\,kHz} & Vevo~\cite{zhang2025vevo} & \textbf{3.88 $\pm$ 0.19} & \textbf{3.43 $\pm$ 0.14} \\
  \cdashline{2-4}
   & LSCodec~\cite{guo2024lscodec} & 3.76 $\pm$ 0.21 & 2.34 $\pm$ 0.19 \\
   & vec2wav2.0~\cite{guo2024vec2wav} & 3.68 $\pm$ 0.20 & 2.63 $\pm$ 0.16 \\
   & \textbf{SDP-Codec-24-S} & \textbf{3.89 $\pm$ 0.18} & \underline{3.20 $\pm$ 0.17} \\
  \cline{1-4}
  \multirow{4}{*}{16\,kHz} & EZ-VC~\cite{joglekar2025ezvc} & 3.19 $\pm$ 0.21 & \textbf{3.51 $\pm$ 0.12} \\
  \cdashline{2-4}
   & BiCodec~\cite{wang2025sparktts} & 3.37 $\pm$ 0.23 & 2.04 $\pm$ 0.19 \\
  & MSRCodec~\cite{li2025msrcodec} & 3.77 $\pm$ 0.20 & 1.93 $\pm$ 0.17 \\
   & \textbf{SDP-Codec-16-L} & \textbf{3.95 $\pm$ 0.18} & \underline{3.46 $\pm$ 0.14} \\
  \bottomrule
  \end{tabular}}
  \vspace{-0.75pt}
\end{table}

\begin{table}[t!]
  \centering
  \footnotesize
  \caption{Ablation study on SDP-Codec.
  w/ $\hat{\mathcal{Z}}$ indicates a model trained with discrete vq-wav2vec features.}
  \label{tab:ablation-pitch-loss}
  {\setlength{\tabcolsep}{3pt}\renewcommand{\arraystretch}{1.1}
  \begin{tabular}{@{}l c c c c c@{}}
  \toprule
  Method & UTMOS &
  \makecell{SECS} &
  \makecell{WER} &
  F0 corr. &
  STOI \\
  \midrule
  \multicolumn{6}{l}{\textbf{Reconstruction}} \\
  \midrule
  SDP-Codec-24-S w/ $\hat{\mathcal{Z}}$ &
  \textbf{4.0659} & 0.9295 & 6.45 & 0.6378 & 0.8541 \\
  SDP-Codec-24-S w/ L2 &
  4.0251 & 0.9326 & 6.45 & 0.6498 & 0.8731 \\
  SDP-Codec-24-S w/o F0 &
  4.0107 & 0.9305 & 5.79 & 0.6367 & 0.8718 \\
  \textbf{SDP-Codec-24-S} &
  \underline{4.0542} & \textbf{0.9353} & \textbf{5.54} & \textbf{0.6522} & \textbf{0.8798} \\
  \midrule
  \multicolumn{6}{l}{\textbf{Zero-shot voice conversion}} \\
  \midrule
  SDP-Codec-24-S w/ $\hat{\mathcal{Z}}$  &
  3.9940 & \textbf{0.8315} & 8.17 & 0.6072 & --- \\
  SDP-Codec-24-S w/ L2 &
  3.9709 & 0.8146 & 8.42 & \textbf{0.6178} & --- \\
  SDP-Codec-24-S w/o F0 &
  3.9670 & 0.8123 & 7.41 & 0.6023 & --- \\
  \textbf{SDP-Codec-24-S} &
  \textbf{4.0055} & 0.8133 & \textbf{7.04} & \textbf{0.6162} & --- \\
  \bottomrule
  \end{tabular}}
  \vspace{-0.25pt}
\end{table}

\section{Results}
\label{sec:results}


\subsection{SDP-Codec-24-S Results}

At 0.45\,kbps, SDP-Codec-24-S matches LSCodec on UTMOS and SECS while improving WER, F0 correlation, and STOI on reconstruction; gains are largest in STOI (0.8798 vs.\ 0.7511) and F0 correlation. It also improves on all zero-shot VC metrics compared to LSCodec.
In subjective VC evaluation, SDP-Codec-24-S achieves the highest NMOS (3.89) among all 24\,kHz systems---matching the larger-scale reference Vevo (3.88)---and the highest SMOS (3.20) among comparable baselines, close to Vevo (3.43).

\subsection{SDP-Codec-16-S and 16-L Results}

At 16\,kHz, SDP-Codec-16-L trades some reconstruction fidelity for substantially stronger zero-shot VC performance than the closest codec-based baselines.
Compared with BiCodec and MSRCodec at similar reported bitrates, it shows lower reconstruction UTMOS and higher WER but achieves the highest VC speaker similarity (SECS 0.8405 vs.\ 0.7577 for BiCodec and 0.7396 for MSRCodec), the highest F0 correlation, and the highest VC UTMOS among codec-based systems, despite BiCodec's higher 0.65\,kbps configuration.
Although EZ-VC attains higher UTMOS, it substantially degrades content fidelity (WER 14.18 vs.\ 4.11), so the gain in naturalness comes at the cost of intelligibility.
Subjective results show an even more favorable trend: despite its higher WER, SDP-Codec-16-L attains the highest NMOS among the evaluated 16\,kHz systems and the highest SMOS among codec-based baselines; EZ-VC attains slightly higher SMOS but much lower NMOS than SDP-Codec-16-L, while BiCodec and MSRCodec remain near 2.0 SMOS, indicating substantially weaker speaker transfer.
In the small-scale setting, SDP-Codec-16-S outperforms XCodec and FocalCodec on SECS and F0 correlation while maintaining competitive reconstruction UTMOS, even at lower bitrates.

\begin{table}[t!]
  \centering
  \footnotesize
  \caption{Speaker probing accuracy of local tokens from various models. (H=HuBERT stream, p=prosody stream, r=residual stream).}
  \label{tab:speaker-probing-mlp}
  \vspace{-0.75pt}
  {\setlength{\tabcolsep}{3pt}\renewcommand{\arraystretch}{1.1}
  \begin{tabular}{@{}l c | l c@{}}
    \toprule
    \multicolumn{2}{c}{\textbf{24\,kHz}} & \multicolumn{2}{c}{\textbf{16\,kHz}} \\
    \cmidrule(lr){1-2} \cmidrule(lr){3-4}
    Method & Acc.\ (\%) $\downarrow$ & Method & Acc.\ (\%) $\downarrow$ \\
    \midrule
    BiCodec & 9.00 & MSRCodec (H) & 15.2 \\
    LSCodec & 15.5 & MSRCodec (H+p) & 39.4 \\
    \textbf{SDP-Codec-24-S} & \textbf{6.87} & MSRCodec (H+p+r) & 46.1 \\
    & & \textbf{SDP-Codec-16-L} & \textbf{4.45} \\
    \bottomrule
  \end{tabular}}
  \vspace{-0.75pt}
\end{table}

\subsection{Ablation Studies}

We ablate the pitch encoder-decoder, the F0 loss formulation, and the use of $\mathcal{Z}$ vs.\ $\hat{\mathcal{Z}}$.
Adding the pitch encoder-decoder with the proposed soft-label loss improves UTMOS, WER and F0 correlation for both reconstruction and voice conversion (Table~\ref{tab:ablation-pitch-loss}), confirming that explicit F0 injection benefits not only prosody but also content fidelity.
In contrast, replacing the soft-label loss with an L2 objective degrades WER to even worse than the no-F0 variant (6.45 vs.\ 5.79), indicating that the benefit of F0 injection depends strongly on the pitch reconstruction objective.

Replacing $\mathcal{Z}$ with $\hat{\mathcal{Z}}$ trades content fidelity for VC speaker similarity (Table~\ref{tab:ablation-pitch-loss}).
Because $\hat{\mathcal{Z}}$ has already been discretized by vq-wav2vec, fine-grained content is lost before SDP-Codec's compact codebook acts, yielding higher WER, despite improving VC speaker similarity modestly.
Because content fidelity remains the main limitation of the current system, we adopt continuous $\mathcal{Z}$ in the final model.

\subsection{Speaker Probing Results}

We probe for speaker information in the local tokens of our main comparison models (BiCodec, LSCodec, and MSRCodec) by training an MLP speaker classifier on the full LibriTTS training set with a per-speaker 9:1 train/test split.
For MSRCodec, we extract each content-related stream (HuBERT, prosody, and residual) and concatenate them frame-wise as input to the MLP.
Table~\ref{tab:speaker-probing-mlp} shows that SDP-Codec yields the lowest speaker-probing accuracy among the evaluated models, which is consistent with reduced speaker leakage in the local token stream; notably, MSRCodec retains substantial speaker information in its prosody and residual streams.

\section{Conclusion}
\label{sec:conclusion}

We present SDP-Codec, a speaker-decoupled low-bitrate neural speech codec trained with a single-stage optimization pipeline.
Across the evaluated 16\,kHz and 24\,kHz settings, SDP-Codec achieves competitive reconstruction quality and strong zero-shot VC performance at comparable reported bitrates, together with the lowest speaker-probing accuracy among the compared systems.
These results suggest that continuous pre-quantization vq-wav2vec features re-quantized through a compact single-codebook bottleneck, explicit F0 injection and the soft-label pitch reconstruction loss together reduce speaker leakage while preserving content and prosodic information.
Future work will focus on further improving content fidelity and applying SDP-Codec tokens to downstream speech language models.

\ifcameraready
\section{Acknowledgments}
This work was supported by the National Research Foundation of Korea (NRF) grant funded by the Korea government (MSIT) (No. RS-2023-00222383).
\fi

\section{Generative AI Use Disclosure}
During the preparation of this manuscript, the authors used generative AI tools for linguistic editing, proofreading, and improving readability. An AI coding assistant was additionally used to help implement and debug the experimental code. All research ideas, methodology, experimental design, and analysis were conceived and carried out by the authors, who reviewed all AI-generated text and code and take full responsibility for the content of this paper.

\bibliographystyle{IEEEtran}
\bibliography{mybib}

\end{document}